\documentclass[aps,prx,reprint,superscriptaddress]{revtex4-1}
\usepackage{graphicx}
\usepackage[]{textcomp}
\usepackage[]{xspace}
\usepackage{dcolumn}
\usepackage{color}
\usepackage{epstopdf}
\begin{document}

\title{Nonlinear optical magnetometry with accessible in situ optical squeezing}

\author{N. Otterstrom}
\affiliation{Department of Physics and Astronomy, Brigham Young University, Provo, UT 84601, USA}
\affiliation{Quantum Information Science Group, Computational Science and Engineering Division, Oak Ridge National Laboratory, Oak Ridge, TN 37830, USA}
\author{R.C. Pooser}
\affiliation{Quantum Information Science Group, Computational Science and Engineering Division, Oak Ridge National Laboratory, Oak Ridge, TN 37830, USA}
\author{B.J. Lawrie}
\affiliation{Quantum Information Science Group, Computational Science and Engineering Division, Oak Ridge National Laboratory, Oak Ridge, TN 37830, USA}

\email{lawriebj@ornl.gov}

\begin{abstract}

We demonstrate compact and accessible squeezed-light magnetometry using four-wave mixing in a single hot rubidium vapor cell. The  strong intrinsic coherence of the four wave mixing process results in nonlinear magneto-optical rotation (NMOR) on each mode of a two mode relative-intensity squeezed state. This framework enables 4.7 dB of quantum noise reduction while the opposing polarization rotation signals of the probe and conjugate fields add to increase the total signal to noise ratio.

\end{abstract}

\maketitle

\section{Introduction}
Ultra-sensitive detection of magnetic fields has crucial biomedical, geological, and astronomical applications \cite{xu2006magnetic,xia2006magnetoencephalography,bick1999hts,dougherty2006identification}. For decades, superconducting quantum interference device (SQUID) magnetometers have dominated much of this field. 
Recently, advances in atomic and optical physics have made optical magnetometry an accurate and cost effective alternative with sensitivity below 1 fT/$\sqrt{Hz}$ \cite{kominis2003subfemtotesla}. Optical magnetometry is fundamentally limited by two noise sources whose sum defines the standard quantum limit: atomic projection noise and photon shot noise. In optimized magnetometers, these two sources are generally found to be comparable \cite{PhysRevA.62.043403,PhysRevLett.95.063004}. The race for greater sensitivity has inspired ongoing research to mitigate both noise sources \cite{PhysRevLett.104.133601,PhysRevLett.105.053601,PhysRevA.86.023803}.

Optical shot noise results from saturation of the Heisenberg uncertainty relation for light, which is given by:
\begin{equation}
\Delta n \Delta \phi = \frac{1}{2}
\label{eq:npr}
\end{equation}
where $\Delta n$ is the uncertainty in photon number and $\Delta\phi$ is the uncertainty in phase.
Squeezed states are quantum states of light that exhibit quantum noise reduction below the shot noise limit in intensity or phase at the expense of uncertainty in the conjugate variable. In recent years, intensity and phase squeezed states have enabled the trace sensing and imaging of signals otherwise obscured by shot noise~\cite{treps_multipixel,treps2002surpassing,Eberle2010,hoff2013,taylor2013}.  Until now, squeezed light magnetometry has been performed by one of two methods: relying on either phased-matched nonlinear crystals~\cite{PhysRevLett.105.053601} or a dual vapor cell arrangement in order to produce a vacuum squeezed state~\cite{PhysRevA.86.023803}. In both cases, the squeezed state generation is performed separately from the magnetometry. Here we present a magnetometer that produces two mode squeezed states from a four wave mixing process while simultaneously performing in situ sub-shot noise magnetometry in a single vapor cell. This technique offers greater quantum noise reduction than any previous squeezed magnetometer in an accessible and compact footprint with no need for an external optical cavity or second cell.

\section{Squeezed State Generation}
\label{sec:Squeezed State Generation}

We generate intensity difference squeezed light by means of four-wave mixing in hot rubidium vapor~\cite{mccormick2007strong,Mccormick2008low,liu2011,qin2012}.
The double lambda system between the hyperfine ground and excited states on the D1 line of ${}^{85}$Rb provides a strong $\chi^{3}$ nonlinearity \cite{shahriar1998generation,lukin1999quantum,lukin1998resonant}. This same system is responsible for the polarization rotation brought on by an external magnetic field in the presence of strong atomic coherence.

Two pump photons generate coherence between the two hyperfine ground states, the strength of which depends on pump power and detuning. The presence of a probe photon redshifted 3044 MHz from the pump field stimulates the coherent emission of a second probe photon and a conjugate photon 3044 MHz blueshifted from the pump in order to conserve both energy and momentum.  Probe and conjugate photons are generated simultaneously and hence share certain quantum correlations, resulting in a two mode relative intensity squeezed state. 

We accomplish this experimentally by superimposing a 300~mW pump field at 795~nm with a weak 20~$\mu$W probe field offset 3044 MHz in frequency in a 1~in.~long rubidium vapor cell held at a temperature of 80~$\pm 1^\circ$~C. The two beam are spatially overlapped in the center of the vapor cell to maximize the effective nonlinear gain. The beams intersect at an angle of approximately 7 mrad, extending the interaction length to the length of the cell and providing an accessible angle to effectively separate probe and conjugate beams from the pump.  This framework has shown considerable potential as a quantum sensing platform for sub-shot noise plasmonics, quantum imaging, and micro-cantilever displacement measurements~\cite{lawrie2013extraordinary,clark2012,Lawrie2013,pooser2014ultrasensitive}.

\begin{figure}[htbp]
\centering
\fbox{\includegraphics[width=\linewidth]{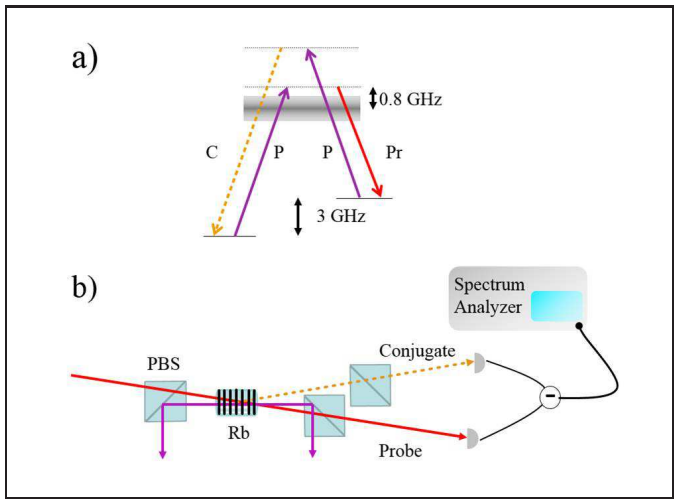}}
\caption{a) Double lambda system for the four-wave mixing process at the D1 (795 nm) line of ${}^{85}$Rb. Two pump photons are absorbed and coherently generate a probe and conjugate photon, conserving both energy and momentum. The blurred line represents the combined hyperfine excited states' narrow splitting compared to the ground states. b) Schematic of four-wave mixing and magnetometry arrangement. The pump field is cross-polarized with the probe and conjugate field in order to easily separate them with polarizing beam splitters after the vapor cell, which also convert polarization rotation to amplitude modulation. External coils around the vapor cell serve to control the temperature, DC magnetic field, and alternating magnetic field.}
\label{fig:setup}
\end{figure}

\section{Squeezed Nonlinear Magneto-Optical Rotation}

The four wave mixing process also provides an opportunity to simultaneously perform subshot noise nonlinear magneto-optical rotation (NMOR) measurements.  The strong, linearly polarized pump field that makes maximal squeezing possible also aligns the rubidium magnetic dipole moments.  
If a magnetic field is applied along the direction of light propagation, then the magnetic moments precess at the Larmor frequency, which is given by:

\begin{equation}
\omega = \frac{egB}{2m},
\label{eq:larmor}
\end{equation}
where $-e$ is the charge of the electron, $g$ is the Land\'{e} factor, $B$ is magnetic field strength, and $m$ is the mass. Hence, by measuring the Larmor frequency we can accurately deduce the magnitude of an applied DC magnetic field~\cite{srinivasan2007nonlinear}. After pumping, the comparatively weak probe and conjugate fields align themselves with the magnetic dipole moments, thus inducing a polarization rotation on both fields.  The relative intensity two mode squeezed state is filtered by two polarizing beam splitters, converting polarization modulation into intensity modulation, and enabling sub-shot noise measurements of polarization rotation.

Traditionally, NMOR magnetometry is performed at very low atomic densities in order to prevent spin relaxation inducing collisions.  Many systems employ a special anti-relaxation coating to preserve coherence by allowing the spins to survive collisions with the cell walls~\cite{budker1998nonlinear,PhysRevA.62.043403,balabas2006magnetometry}. If, however, the rate of spin exchange is higher than the rate at which the moments precess, such collisions cause limited decoherence~\cite{PhysRevLett.89.130801}. Our experiment uses a pump intensity two to three orders of magnitude larger than previous squeezed light magnetometers in order to provide appreciable nonlinear gain. The detuning, $\Delta$, is similar to other experiments at approximately 1~GHz, meaning that the Rabi frequency, given by $\sqrt{\Omega_0^2+\Delta^2}$, with $\Omega_0= \left<d\right>E_0/\hbar$ (where $d$ is the electric dipole transition element and $E_0$ is the magnitude of the electric field), in our experiment is large compared to other magnetometers. This results in two to three orders of magnitude larger atomic coherence than other squeezed magnetometers. We operate at high atomic densities and without parrafin coatings on the cell surfaces, maximizing optical transmission. The Rabi frequency is large enough to guarantee a long-term average coherence compared to the amount of time the probe photons take to traverse the optically pumped Rb atoms. This can be verified empirically by noting a large nonlinear gain, on the order of 12.6 for this experiment. With atomic densities of about $1.36 \times 10^{12}$ our magnetometer falls into a regime where spin coherence is preserved for NMOR measurements without the need for a paraffin coated cell.

\begin{figure}[ht]
\centering
\fbox{\includegraphics[width=\linewidth]{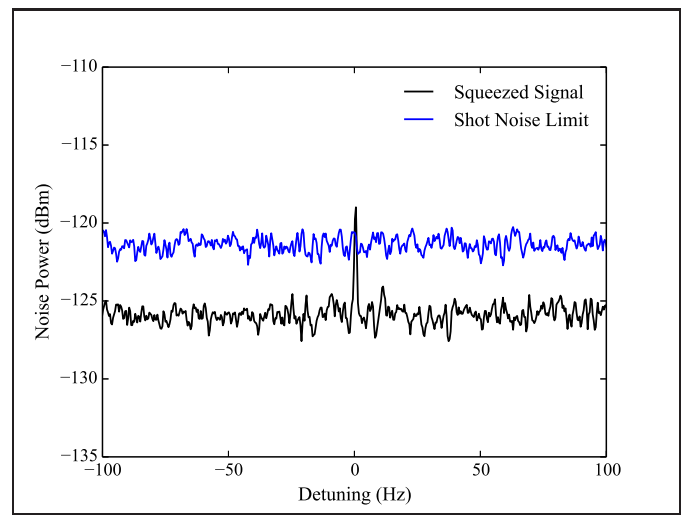}}
\caption{Noise spectrum of NMOR signal due to an alternating field of 37.5 pT at 700 kHz with 4.5$\pm{0.1}$ dB of quantum noise reduction (RBW = 1 Hz, VBW = 100 Hz).   Quantum noise reduction enables a minimal resolvable signal below the shot noise limit.}
\label{fig:subshotnoise}
\end{figure}

Frequency modulated nonlinear magneto-optical rotation~(FM NMOR)~\cite{RevModPhys.74.1153} measures weak DC magnetic fields at RF frequencies dominated by photon shot noise: a regime that enables quantum noise reduction with squeezed states of light. This method involves precise measurement of the Larmor frequncy, and is fundamentally limited by the line-width and the signal to noise ratio of the resonance~\cite{PhysRevLett.95.063004}. Although insufficent magnetic shielding prevents high sensitivity DC field measurements in this experiment, the introduction of a weak AC magnetic field enables squeezed NMOR analogous to the FM NMOR framework~\cite{PhysRevLett.105.053601,RevModPhys.74.1153}. The DC sensitivity could be scaled to the sub-fT regime by incorporating this approach to a traditional FM NMOR geometry.

\begin{figure}[ht]
\centering
\fbox{\includegraphics[width=\linewidth]{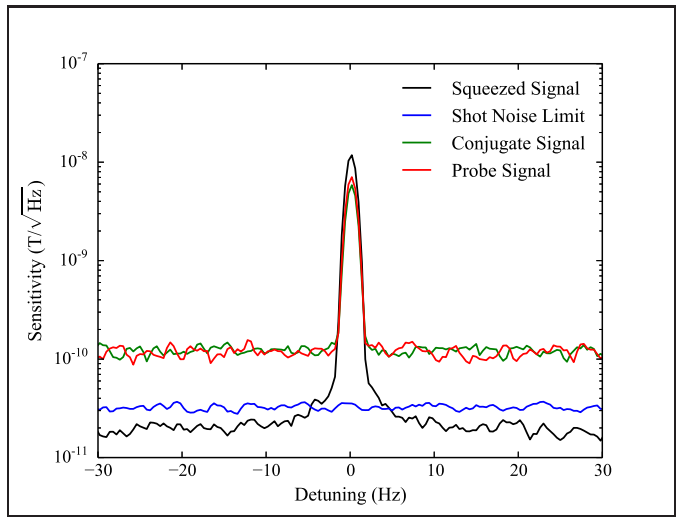}}
\caption{Noise spectrum of squeezed NMOR signal due to an alternating field of 11.8 nT peak-to-peak at 700 kHz (RBW = 1 Hz, VBW = 100 Hz). All signals are averaged over 100 measurements. The red and green curves show the probe and conjugate signals individually, confirming that both probe and conjugate fields experience NMOR. As one channel is blocked to measure the other, the noise floors on green and red curves are dominated by classical noise. The black curve represents the NMOR signal when subtracting probe and conjugate fields on the detector. We orient the two polarizing beam splitters such that the NMOR signals on each channel add while subtracting classical and quantum noise. This measurement corresponds to 4.7$\pm{0.1}$ dB of quantum noise reduction, improving the AC sensitivity of the magnetometer from 33.2 pT/$\sqrt{Hz}$ to 19.3 pT/$\sqrt{Hz}$.}
\label{fig:conjprobe}
\end{figure}

\begin{figure}[ht]
\centering
\fbox{\includegraphics[width=\linewidth]{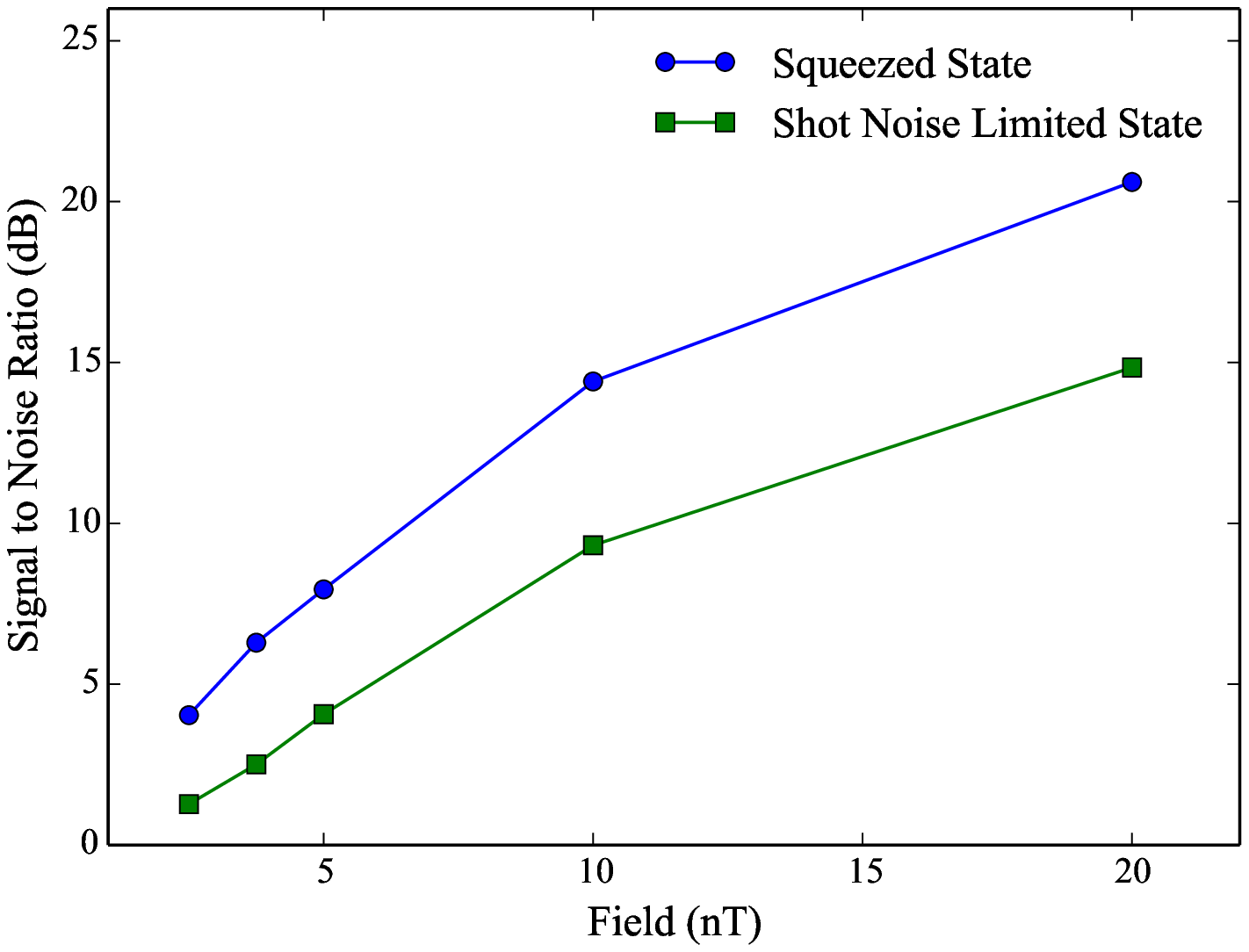}}
\caption{Signal to noise ratio (SNR) of squeezed and shot noise limited signals at varying AC magnetic field strengths. The Spectrum analyzer was set to RBW = 10 KHz and VBW = 100 Hz, and the signal was averaged over 10 samples. The uncertainty for each measurement is less than $\pm{0.1}$ dB. }
\label{fig:snr}
\end{figure}

For the atomic density and detuning used in this experiment, the probe transmission through the cell is 86\%. The transverse DC magnetic field within the vapor cell is reduced to less than 10 $\mu$T by varying the DC current on an exterior solenoid. Measurements from a spectrum analyzer are used to determine the quantum enhanced sensitivity of the magnetometer. We apply a weak sinusoidal magnetic field with a solenoid about the vapor cell's exterior.  AC magnetic field sensitivity remained unchanged in the 300-900 KHz range, and we report our measurements at 700 KHz.
The optical magnetometer was calibrated using a linear magnetic field sensor, but the limiting factor in sensitivity for this experiment arose from the spectrum analyzer, which has a minimum RBW of 1 Hz.

Figure 2 shows a NMOR measurement at 700~kHz that would otherwise have been obscured by the shot noise. This modulation is unresolvable on the individual probe and conjugate channels, but a difference measurement enables subshot noise NMOR measurements and enhances the SNR over a large range of field strengths as shown in Figs.~2-4. Figure 3 demonstrates that quantum noise reduction improves the SNR of large signals and that both probe and conjugate fields experience NMOR, despite the conjugate field being far from the atomic resonance.  This phenomenon is due to the fact that the stimulating probe photon coherently imparts its polarization to the entangled probe and conjugate photon pair. By independently orienting the two polarizing beam splitters after the vapor cell, we reflect the respective polarizations about the y-axis. Hence, the two beam splitters act as filters of opposite sign, allowing the NMOR signals on each channel to add while subtracting classical and quantum noise.  At this field strength the measured signal has an instrument-limited line-width of about 2 Hz (FWHM). The minimum resolvable signal of the magnetometer~\cite{PhysRevA.86.023803}, is enhanced by 4.7$\pm{0.1}$ dB from 33.2 pT/$\sqrt{Hz}$ to 19.3 pT/$\sqrt{Hz}$, state of the art for squeezed light magnetometers~\cite{PhysRevLett.105.053601,PhysRevA.86.023803}.

\section{Conclusion}

We have demonstrated an alkali atomic magnetometer with 4.7~$\pm~0.1$~dB of in situ optical squeezing.  This magnetometer benefits from quantum noise reduction in an accessible and compact footprint, enabling sub-shot noise measurements of AC and potentially DC magnetic fields by adapting our method to FM NMOR magnetometry.

\section*{Funding Information}
This work was performed at Oak Ridge National Laboratory, operated by UT-Battelle for the U.S.Department of energy under Contract No. DE-AC05-00OR22725, and was supported in part by the U.S. Department of Energy, Office of Science, Office of Workforce Development for Teachers and Scientists (WDTS) under the SULI program. B.~L.~and R.~C.~P~ acknowledge support from the lab directed research and development program.




\end{document}